# A Unified Multi-scale and Multi-task Learning Framework for Driver Behaviors Reasoning

Yang Xing [a], Chen Lv [a], Dongpu Cao [b], Efstathios Velenis [c]

[a] School of Mechanical and Aerospace Engineering, Nanyang Technological University, 639798, Singapore.
[b] Mechatronics Engineering with the University of Waterloo, 200 University Avenue West Waterloo, ON, N2L3G1, Canada.
[c] Advanced Vehicle Engineering Centre, Cranfield University, Bedford, MK43 0AL, UK.

*Abstract*—**Mutual understanding between driver and vehicle is critically important to the design of intelligent vehicles and customized interaction interface. In this study, a unified driver behavior reasoning system toward multi-scale and multi-tasks behavior recognition is proposed. Specifically, a multi-scale driver behavior recognition system is designed to recognize both the driver's physical and mental states based on a deep encoder-decoder framework. This system can jointly recognize three driver behaviors with different time scales based on the shared encoder network. Driver body postures and mental behaviors include intention and emotion are studied and identified. The encoder network is designed based on a deep convolutional neural network (CNN), and several decoders for different driver states estimation are proposed with fully connected (FC) and long short-term memory (LSTM) based recurrent neural networks (RNN). The joint feature learning with the CNN encoder increases the computational efficiency and feature diversity, while the customized decoders enable an efficient multi-tasks inference. The proposed framework can be used as a solution to exploit the relationship between different driver states, and it is found that when drivers generate lane change intentions, their emotions usually keep neutral state and more focus on the task. Two naturalistic datasets are used to investigate the model performance, which is a local highway dataset, namely, CranData and one public dataset from Brain4Cars. The testing results on these two datasets show accurate performance and outperform existing methods on driver postures, intention, and emotion recognition.**

## I. INTRODUCTION

### A. Motivation

Driving is a complex task for the human driver, which usually require various adjustment in the physical, emotional, cognitive, and psychological aspects. Drivers should efficiently interact with other road entities based on their context perception, decision-making, and control action to the vehicle. With the development of intelligent and autonomous vehicles, a common agreement has been made that human drivers can be a good teacher to the intelligent agents in general [1]. Hence, a comprehensive analysis of human drivers and learning driving patterns and styles from drivers can benefit the human-centered intelligent system design [2]. As drivers are sharing their vehicle control authorities with the intelligent units, conflict can occur if the driver and vehicle cannot well understand with each other. Therefore, it is critical to design efficient driver-vehicle collaboration and shared control strategies based on driving behavior prediction and driver mental state inference [3,4]. Moreover, mutual-understanding enabled intelligent vehicles are much easier to be accepted by the public as human drivers/passengers can feel they are well-considered and understood so that to be confident with the intelligent vehicles [5].

Driver behaviors recognition is a wide scope of research and has been widely studied in the past two decades. Driver behaviors can be divided into multiple scales, from seconds-level activities and cognitive process to hours or days-level of driving styles, skills, and habits. Among these, a large number of studies focus on the in-vehicle driver states estimation, such as driver attention, driver intention, emotion, and fatigue, etc., which are essential to driving safety issues [6,7]. In the past, it is not well studied how to understand multi-scales driver behaviors uniformly and how these behaviors influence each other. Existing studies mainly focus on a specific topic to understand and model the driver from a single aspect. As a result, it would be challenging to integrate so many different functional modules with varying techniques into a mutual understanding system on intelligent vehicles.

Therefore, in this study, a driver behavior reasoning system is proposed based on a unified framework to increase the scalability, accuracy, and efficiency of the system. The unified driver behavior recognition system can estimate multi-scale driver behaviors at both a physical-level and mental-levels. The definition of multi-scale driver behaviors can be found in [8], where three different behaviors were identified, namely, strategical behavior, tactical behavior, and operational behavior. The time scales for the three kinds of behaviors are in descending order. The driver's physical behaviors, such as mirror checking and facial expressions, are at the operational level, which has fast and distinctive dynamics that usually can be recognized based on a single image. While the mental cognitive process, such as the intention and emotion, can have longer time-scale and vibrant temporal patterns, which need to be inferred based on sequential streams and time series models. It is needed to capture both the spatial and temporal patterns from the driver monitoring system in order to jointly estimates different driver states. Therefore, in this study, an encoder-decoder deep learning architecture is adopted for multi-scale and multi-tasks driver reasoning. The unified framework uses a shared deep CNN model as the encoder to extract in-vehicle spatial features of the human driver, and several LSTM-RNN models are trained to capture the temporal patterns of the mental cognitive processes. The proposed framework is separable and scalable and can be extended by introducing more light-weight decoders for driver state estimation.

*B. Literature Review*

Driver intention inference is a process to anticipate the near-future driving maneuvers based on the reasoning of driver physical behaviors, traffic context, and vehicle states. To infer the driver's intention, vision-based methods are the most efficient categories as the driver's intention is closely related to driver behavior recognition, such as mirror checking detection before the maneuver [9]. In [10], the authors developed a hidden Markov model (HMM) to model the probabilistic transitions of the mental states and predict the lane change intent on a driving simulator. Then, in [11], a sparse Bayesian learning network was developed to predict the lane change intent with naturalistic data. In [12], the impact of head movement and gaze movement on the intention inference was analyzed. It was found that the head movement before the maneuver was the most important clue for the intention estimation. At the same time, eye gaze signals did not significantly improve the model performance. In [13], a lane change intention prediction system considering different driving styles was proposed based on the Bayesian network (BN) and Gaussian mixture model (GMM). It showed that by integrating driving styles and traffic context, the lane change intent could be predicted 4.5 s ahead of the maneuver with 78.2% accuracy. As driver intention inference requires the modeling of temporal driver behavior dynamics; recently, some researchers applied deep learning models to solve the task. In [14], a driver intention and path prediction system for urban intersection driving behavior modeling was proposed based on the LSTM-RNN. Similarly, in [15], an LSTM-RNN framework was applied to anticipate the lane change and turn maneuvers based on the multi-modal data. The model can predict the lane change maneuver 3.42 s before it happens with a precision and recall of 88% and 86%, respectively. Although reasonable results have achieved in the past, most of the existing methods rely on a hand-craft feature engineering to extract features, which lacks objective evaluation and normally requires extra modules and functions such as head pose and eye gaze estimation to extract the features.

Regarding driver emotion analysis, despite the neutral state, seven universal human emotions can be reflected by the facial movements, which are anger, fear, disgust, sadness, surprise, contempt, and happiness [16]. It was shown that both the positive emotional stimuli (happy) and negative emotional stimuli (fear, anger, etc.) could worsen the driving performance on lane-keeping, traffic role violations, and aggressive driving [17]. Therefore, it is necessary to recognize the driver's emotions to provide proper alarms, assistant, and shared controls for driving safety issues. Driver emotion recognition usually can be detected with the multi-modal signal, which can be grouped into three primary information, which are facial expression, speech, and electroencephalogram (EEG) and other physiological signals [18]-[21]. For example, in [22], an emotion recognition system with the inner cabin voice signal was proposed based on the prosodic and spectral features extraction. The authors in [23] argue that the visual and audio signals can be less effective in the advanced driver assistance system (ADAS) due to the facial expression can be faked, and the audio signals may not always available. Hence, a subject independent emotion recognition system based on the electrodermal activity (EDA), skin temperature (ST), an electrocardiogram (ECG) signals were proposed. In fact, the facial expression is the most informative interpersonal communication channel that carries about two-thirds of the total communication information [24]. Methodologies for facial emotion recognition (FER) can be roughly divided into two categorical, which are conventional methods based on feature extraction [25,26] and end-to-end deep learning-based methods [27]-[31]. The deep learning-based FER has achieved state-of-are results on many public datasets. However, one of the drawbacks of these systems is the lack of temporal dependency analysis during a long-term driving process. Specifically, the emotional states are not transient and should not be classified based on a single image during the driving task. Within a certain period, the variation of the driver's emotion is not that fast, and sometimes emotion recognition needs to be made based on the previous context and driver states.

In this study, a combined CNN-RNN model is developed to recognize these driver state uniformly. The CNN-RNN network has become a powerful model for sequential image processing [32,33]. In [34], A long-term recurrent convolutional network (LRCN) was proposed. The LRCN can map variable-length input to variable-length output and capture complex temporal dynamics. In [35], a driver hand gesture prediction model was proposed based on the combined LSTM and CNN network to detect the low latency gestures. In [36], a convolutional LSTM (ConvLSTM) was built for precipitation nowcasting by extending the fully connected LSTM to a convolutional structure. In this study, a multi-task framework is integrated into the CNN-RNN network to process the multi-scale driver behavior recognition task. Multi-task learning aims to leverage common representation and useful information in several related tasks to enhance the generalization performance on multiple tasks [37][38]. Currently, many studies show the multi-task learning paradigm can achieve compatible and even better results compared with single-task learning. In [39], an encoder-decoder multi-task learning network for road segmentation, object detection, and classification was designed. In [40], a partially shared CNN architecture was developed for simultaneous gaze point and gaze direction estimation. The multi-task learning paradigm has also been utilized in many perceptions, translation, and classification tasks [41,42]. However, very few studies have analyzed the effectiveness of utilizing such a framework for the improvement of driver behavior reasoning, especially focus on the learning of multi-scale driver behaviors towards a unified and intelligent mutual-understanding scheme. To our best knowledge, this study is the first one to exploit a unified multi-task and multi-scale driver understanding model that can recognize both physical behaviors and mental activities for human drivers.

### C. Contribution

As aforementioned, driver behavior recognition has been widely studied. However, very few studies exploit the common patterns and representations between different behaviors. In this study, a unified driver behavior reasoning framework is designed to estimate the multi-scale driver behaviors. The contribution of this study can be summarized as follows.

First, an encoder-decoder based unified network structure is proposed for multi-scale driver behavior learning and reasoning. The network is flexible and scalable that can be expanded to enriching the representation in the encoder part and designing a proper state estimation module in the decoder part. Second, based on the proposed network, it is proved that the higher-level unobservable driver mental states such as intention and emotion can be jointly learned and inferred based on the lower-level observable driver pose and facial representations. The unified network is evaluated based on the different datasets and achieved state-of-the-art results compared with existing methods. Last, different driver states are jointly analyzed based on the proposed network. This analysis can benefit the intelligent and autonomous vehicles towards a better mutual understanding system.

### D. Paper Organization

The remainder of this paper is organized as follows. Section II introduces the high-level framework of the unified driver behavior reasoning and the experimental setup. In section III, the model structures are highlighted, and the training process is clarified. The experiment results and model evaluation for the different tasks with the different datasets are represented in Section IV. In Section V, the discussion on the proposed model and future works is proposed. Last, the study is concluded in Section VI.

## II. SYSTEM OVERVIEW

In this section, the high-level structure of the proposed unified driver behavior reasoning system will be described. Specifically, three key aspects are introduced, which are naturalistic data collection and processing, CNN based driver physical behavior model construction, and temporal sequence processing for driver mental state inference based on LSTM-RNN.

### A. High-Level System Architecture

The overall framework is shown in Fig. 1 below. Three major components are designed, namely, data collection and processing, driver physical behavior recognition, and driver mental state inference. First, the naturalistic driving behavior on the highway is collected. A data collection and synchronization system were designed to capture both driver behaviors information and traffic context. Three cameras are implemented inside the vehicle cabin to collect driver front face images, hand movements, and traffic context. A detailed experiment testbed setup is described in the next part.

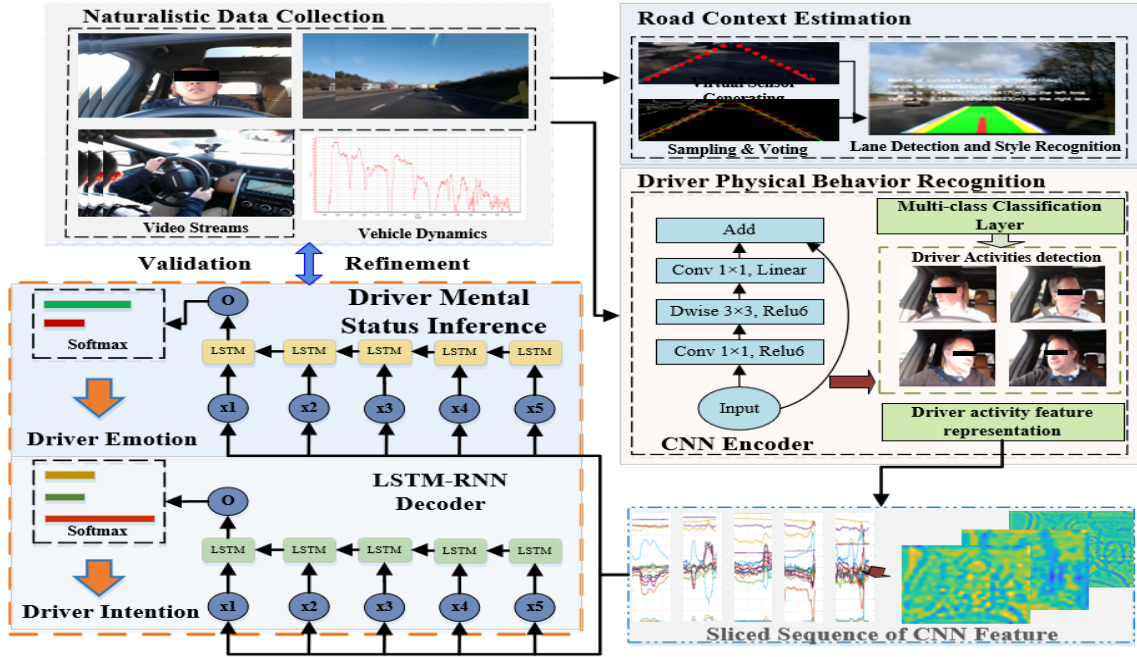

Fig. 1. Illustration of the high-level architecture of the unified multi-task driver behavior reasoning system.

Once the naturalistic driving data are collected and synchronized, these data need further processing to train and evaluate the system. The road information from the front-looking cameras will be used to extract the lane positions, lane styles, and road curvature. The road context information enables a more reasonable driver lane change intention inference as a solid lane means the driver cannot make a lane change in that direction. The second module of the proposed system is a CNN based encoder for spatial feature learning and extraction. In this part, two different pre-trained CNN models, namely, MobileNet V2 and Inception-Resnet V2 will be adopted and compared [43,44]. Based on the CNN encoder, high-level representation for driver physical behavior related features can be learned and extracted. Each frame of the sequence will be fed into the CNN encoder, and the output of each frame will be concatenated into a sequential feature tensor for further time-series model construction.

The driver's mental state inference module contains two separate parts, which are driver emotion recognition and driver intention inference. Driver mental states differ in physical behaviors from two aspects. First, mental states like driver intention are non-observable and can only be inferred based on driver physical behaviors such as head pose, and body movement. Second, these cognitive processes are usually depending on many aspects, such as self-motivation and traffic context stimuli; these processes are dynamic and require time-series modeling to capture the temporal pattern between previous and current behaviors. The decoder for driver physical behavior recognition part is simply fully-connected layers, while the LSTM-RNN based models are adopted for driver mental states inference. Driver emotion variation is viewed as a cognitive process in this study as driver emotion can be hardly changed very quickly within a short period. The encoder-decoder multi-task learning represents a simplified version of the human cognitive process. The CNN encoder plays a similar role to the human visual system, which mainly focuses on perception and feature extraction. On the other hand, the multiple RNN decoders mimic the brain reasoning parts to generate different inference results for different tasks.

*B. Experimental Data Processing*

In this study, two naturalistic datasets are used to evaluate the proposed system. The first one is the collected highway dataset around the Cranfield area in the UK (known as CranData). The second one is a public dataset, namely, Brian4Cars dataset, which is available at http://www.brain4cars.com [45]. The CranData was collected based on three cameras and one VBOX GPS logger for vehicle speed and heading measurement. The three cameras are mounted inside the vehicle for the driver's head and upper body monitoring, hand movement recording, and traffic context recording, respectively. The three video streams are synchronized and recorded at a frequency of 25 fps with $640 \times 480$ resolution. The default sampling rate of the VBOX data logger is 20 HZ. Three male drivers with varying ages and experiences were involved in the data collection. All of them were asked to drive as usual without telling them the real objective of the experiment. Each driver drove the vehicle on the highway for about one hour. The Brain4Cars dataset contains both inside and outside videos, which were sampled at 25 fps and 30 fps, respectively. The data consisted of 1180 miles of freeway and city driving and was collected from 10 drivers. The data also annotated the number of lanes on the road and the current lane. Five driving maneuvers were recorded, which are driving straight, lane change left/right, and turn left/right. For comparison reason, only the first three maneuvers are used, and the turn maneuvers are not studied in this study.

## III. METHODOLOGIES

In this section, the multi-task learning model is described. The encoder part of the model is a CNN model that extracts the temporal-spatial abstract features from the driver behavior image sequence. Then, different decoder networks are trained separately to personalized estimate the specific task.

*A. Model Architecture*

The overall model architecture is shown in Fig.2 below. The model follows an encoder-decoder CNN-RNN structure so that the shared abstract spatial driver behavior features can be used for multi-task learning and prediction. A CNN model is used as an encoder for sequential feature extraction. Then, in the middle layer, a feature fusion module is developed before feeding the sequential features into the decoding part. The feature fusion layer makes the model even flexible and enables scalable implementation as the model can be easily integrated with other existing modules. The modularity design method is an efficient fashion to integrate different features together for comprehensive driver states estimation. For instance, the road context information can be integrated into the driver behavior feature tensor through the middle feature fusion layer to contribute a more precise driver lane change intention inference. Also, the onboard speech recording system can extract the speech and audio information between the driver and passengers so that the speech features can be involved in the driver's emotion recognition process. Once the features from different modules are fused, the temporal feature sequence can be fed into the multi-task decoders. The encoder-decoder multi-task learning framework represents a simplified version of the human cognitive process. The CNN encoder plays a similar role to the human visual system. The task of the encoder part is to extract the representative features that can reflect driver outer physical behaviors and states.

The encoder part consists of several convolutional layers and pooling layers of a deep classification network to construct a strong vision-based system to extract important spatial features from the video sequence. Specifically, two state-of-the-art pre-trained deep CNN networks, namely, MobileNet V2 and Inception-Restnet V2 are adopted. The structure of the two pre-trained networks is very different from each other. Although both of them achieved state-of-the-art results on image classification, object detection, and segmentation, etc., MobileNet is a light but an efficient network that is designed for embedded and mobile computing, which only contains 155 layers. On the contrary, the Inception-Resnet V2 is a much deeper network that contains 825 layers, and the feature representation capability can be more powerful than the MobileNet. The MobileNet solution can be used to investigate the real-time performance of the proposed framework on the embedded computational platform. In contrast, the Inception-Resnet V2 can be used to exploit the maximum model capability. The last fully connected and softmax layers of the two encoder networks are discarded and replaced with two extra fully connected layers for driver physical behavior recognition and feature extraction.

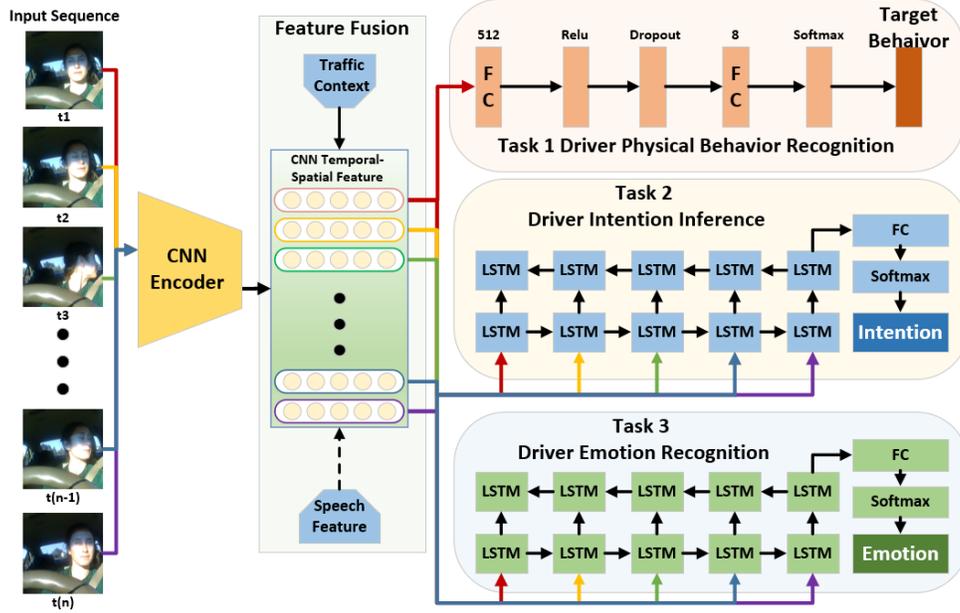

Fig. 2. Multi-task learning model framework for driver behavior reasoning. A pre-trained deep CNN is fine-tuned based on driver behavioral data and used as the CNN encoder. Three decoders are designed to estimate the multi-scale driver behaviors on both physical and mental levels. The middle feature fusion model is designed to extend the model ability by integrating more features from other modules. The dashed line means the speech feature is not used in this study but can be involved in the future.

As shown in Fig. 2, the decoding part will be responsible for three different tasks, which is driver physical behavior recognition, driver intention inference, and driver emotion recognition, respectively. The driver's physical behavior recognition task jointly estimates the driver's mirror checking behaviors and the driver's facial expression. There are four mirror checking behaviors used in this study, which are front-facing, rear mirror checking, left mirror checking, and right mirror checking. Besides, there are two facial expressions for each image, which are the normal and emotional expression. Each mirror checking behaviors map two possible facial expressions so that eight different physical behaviors are identified in total. The final output of the FC decoder is the estimated behavior for the single image at each step, and no temporal patterns are required.

The two driver mental states inference tasks are similar to each other and require the temporal patterns for real-time inference. The driver intention inference module is designed to infer the lane change intention based on the sequential input tensor from the encoder. Three intentions, namely, lane change right, lane change left, and lane-keeping, are classified. The emotion recognition task focuses on the detection of neutral and emotional states based on the sequential inputs. The bidirectional LSTM-RNN (BiRNN) architecture is applied as BiRNN can capture more forward and backward dependency patterns of the sequential feature tensor. The detailed BiRNN model structure for these two models is shown in Fig. 2. Bidirectional LSTM layers with 150 hidden units in each layer are implemented.

The BiRNN can be represented as follows, the forward and backward hidden states of the current time are the function of previous states, and the input tensor [46]. The final output is a function of the forward states and backward states.

$$s_{ft} = f(W_1 x_t + W_2 s_{ft-1} + b_x) \tag{1}$$

$$s_{bt} = f(W_3 x_t + W_5 s_{bt-1} + b_x) \tag{2}$$

$$o_t = f(W_4 s_{ft} + W_6 s_{bt} + b_o) \tag{3}$$

The LSTM cell introduce three gates to forget and update the historical information, which are forget gates (), input gates (), and output gates ().

$$f_t = \sigma(U_f x_t + W_f s_{t-1} + b_f) \tag{4}$$

$$i_t = \sigma(U_i x_t + W_i s_{t-1} + b_i) \tag{5}$$

$$o_t = \sigma(U_o x_t + W_o s_{t-1} + b_o) \tag{6}$$

The value $\widetilde{c_t}$ is the candidate cell state which can be represented as:

$$\widetilde{c_t} = tanh(U_c x_t + W_c s_{t-1} + b_c) \tag{7}$$

The $c_t$ in the center is the internal memory cell state of the LSTM unit, which is the combination of previous $c_{t-1}$ and current candidate states.

$$C_t = f_t * c_{t-1} + i_t * \widetilde{c_t} \tag{8}$$

Last, the layer output is the products of the cell state $C_t$ and the candidate output from the output gate.

$$s_t = o_t * tanh(C_t) \tag{9}$$

Detailed explanation for LSTM cell can be found in [47].

## B. Model Training and Testing

The training strategy for the multi-task driver behavior reasoning model follows a fine-tuning process. The encoder CNN model is initialized with the weights trained on the ImageNet [48]. Then, the last FC layers are discarded and replaced with new FC layers and classification layers for driver physical behavior classification. Driver physical behaviors and facial expressions are low-level driver activities, which are the outer reflection of the mental states. Hence, the encoder trained with physical behavior data can be used as an informative representation of the mental states. A $10^{-4}$ initial learning rate is applied for the convolutional layers of the CNN encoder to decrease the learning rate and maintain the feature extraction power. The weight learning rate and bias learning rate of the new FC layers are selected as 20. The mini-batch size for the MobileNet is 32, and 15 for the Inception-ResNet dues to the memory capability. There are 47111 samples in the CranData and 36760 samples in the Brain4Cars dataset. 80% of the data are split into the training set, while the rest are used for testing. Each model is trained with three epochs.

The lane change intention inference and emotion recognition decoders are trained with sequential inputs. The input tensors are generated by the encoder and the feature fusion layer. The activation from the second last FC layers (with 512 neurons) is used for feature extraction. The sequence length is varied according to the different length of the input sequence in the CranData. Each record in the Brain4Cars dataset is a sequence with 150 images for six seconds driving data. An initial learning rate of 0.1 with a 0.5 decay rate for every 50 epochs is applied. The mini-batch size is 32, and the max training epoch is 500. There are 201 samples within the CranData, and 244 samples are collected from the Brain4Cars dataset. The cross-entropy loss function is used, and the models are optimized with Adam optimizer [49]. The overall testing rate achieved around 25 fps on a low-cost Nvidia GPU (MX150), which can satisfy the real-world application requirement.

Unlike some multi-task learning methods which jointly training the encoder and decoder part with a summed or weighted summed cost function [38,39], in this study, the decoder and encoder parts are trained separately. The in-vehicle driver monitoring systems usually need modularized design. The vision-based system is not the only on-board system for driver states reasoning on intelligent and autonomous vehicles. For example, there can be other driver assistance systems such as radio and human-machine-interaction (HMI). Separately training the encoder and decoders is more flexible, as more sensors and features can be easily integrated into the feature fusion layer, and only the decoders need to be updated if the input tensor changed. Hence, the CNN encoder in this study is merely used as a task-oriented feature extraction module.

## IV. EXPERIMENTAL RESULTS

In this section, the experimental results and model evaluation will be illustrated for the three different tasks. The classification results for the eight basic physical behaviors are first proposed to show the feature representation capability of the encoder CNN. Then, the time-series classification for intention inference and emotion recognition will be evaluated. Each of the tasks will be evaluated with two datasets, respectively.

### A. Driver Physical Behaviors Classification

The driver physical behavior classification task is first evaluated as precise driver observable behavior representation is the fundamental part of mental state modeling. The two pre-trained model performance on the CranData and Brain4Cars dataset are assessed separately. This part focusses on the evaluation of the two selected models based on the two datasets. The visualization of the learned feature representation of the two models is shown in Fig. 3. Two behaviors, namely, normal emotional (happy in this case) driving and neutral right mirror checking behaviors are investigated. The occlusion sensitivity method [50] is used to represents the model representation capability. It is shown in Fig. 3 that all of the two models learn a good representation of the combined states of driver behaviors and particular focus on the facial area to estimate the head pose and facial expression jointly.

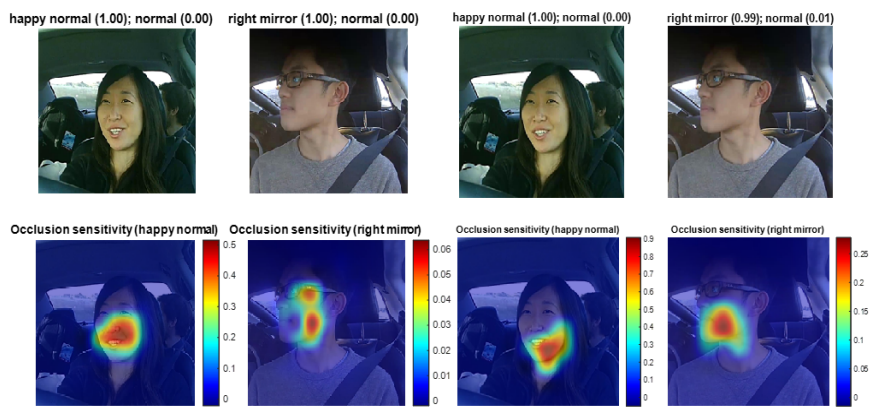

Fig. 3. Model representation visualization results in the normal emotional driving, and neutral right mirror checking behaviors, respectively. The left four results are generated by MobileNetv2, and the right four are generated by Inception-ResnetV2. The top row indicates the behavior classification results, and the bottom row indicates the occlusion results.

The classification results for the eight driver behaviors are shown in Fig 4. below. It should be mentioned that in the Brain4Cars dataset, there are quite a few rear mirrors checking case, and it is hard to find emotional rear mirror checking, hence, in this data, only seven behaviors are investigated and classified. As shown in Fig 4, both of the two models can achieve more than 90% classifications on the two datasets. The Inception-ResNet performs slightly better on the CranData, while, the ModelNet achieved a more precise result on the Brain4Cars dataset. Both of the light MobileNet and heavy volume Inception-ResNet can learn the representative features of the driver's physical behaviors and generate consistent classification results on the two datasets.

Fig. 4. Confusion matrix of the behavior classification based on MobileNetV2 and Inception-ResNetV2 on the CranData and Brain4Car datasets.

### B. Driver Intention Inference

In this part, driver intention inference results are compared between multiple baseline algorithms on the two datasets. The baselines methods are first introduced as follows. As emotional recognition follows a similar time-series modeling process, the same baseline methods will be used in the next part. The CNN-RNN based driver intention model will be compared with the existing methods, which use head pose and eye gaze tracking for driver feature extraction following with machine learning methods for classification. For the CranData, the in-vehicle driver head poses and out-vehicle traffic context information will be concatenated. Specifically, the driver head pose and eye gaze feature will be detected based on a Conditional Local Neural Fields (CLNF) approach [51]. The inside feature vector at each time constant can be formed as follows.

$$I_t = [G_r \ G_l \ G_a \ H_t \ H_r] \tag{10}$$

where $G_a$, is the 2D gaze angle in $x$ and $y$ coordinate, $G_r$, $G_l$, $H_t$, and $H_r$ are the 3D gaze direction for each eye, head pose translation vector, and head pose direction vector, respectively. This gives a 14-dimensional inside the vehicle feature vector. As the CranData was collected on a highway, the line styles of the ego-lane are used. Two vehicle dynamic signals, namely the vehicle speed ($V$), and heading angle ($H$) are collected using the VOBX. The total feature vector for the outside traffic context and vehicular dynamics can be formed as a four-dimensional vector.

$$O_t = [L_r \ L_l \ V \ H] \tag{11}$$

where $L_r$ and $L_l$ are the lane style for right and left lane and $L \in [0, 1]$ represent the two different lane styles (solid and dash). The in-vehicle feature $I_t$ and out-vehicle feature $O_t$ will be concatenated to form an 18-dimensional feature vector.

According to the developed feature extraction method, several baselines are designed as follows.

1. **Support Vector Machine (SVM).** SVM is a discriminative classifier, which was used in the past for driver intention detection [11,12]. As SVM cannot process the sequential data directly, a statistical feature vector will be used to calculate the maximum, minimum, mean, and standard deviation (STD) of the head pose features. Hence, the 18D feature vector for each step will be expanded to a 72D feature vector for the whole sequence.

2. **Hidden Markov Model (HMM).** HMM is a generative classifier that assigns a probabilistic graph for each class [10,15]. A fixed-size vector with a length of 150 was used, which carries six seconds inside and outside-vehicle information.

3. **LSTM based RNN.** The LSTM-RNN based method is adopted for intention inference, which is based on the hand-crafted feature vector (**HF-LSTM**) [45].

4. **CNN-RNN encoder-decoder models (CRNN).** The CNN-RNN model is the model shown in Fig. 2, which use CNN as encoder and Bi-directional LSTM-RNN for the decoder. Also, two different encoders, which are noted as **CNN-RNN-M** (MobileNetv2) and **CNN-RNN-IR** (InceptionResnetV2), are evaluated respectively.

To quantitively evaluate the model performance, four performance indexes, namely, precision, recall, F1 scores, and the general average precision, are adopted to assess the model performance on each maneuver (lane change right, lane change left, and lane-keeping). Four metrics are statistically calculated for each maneuver, which is the true positive ($T_p$, the model correct detects this maneuver), true negative ($T_N$, the mode correct detects the other maneuvers), false positive ($F_p$, the model detects the other maneuvers as target one), and false negative ($F_N$, the mode detects the target maneuver as other maneuvers).

According to the four-performance index, the Precision ($P_r$) can be calculated as:

$$\mathrm{Pr} = \frac{T_P}{T_P + F_P} \tag{12}$$

The Recall ($R_e$) is calculated as:

$$\mathrm{Re} = \frac{T_p}{T_p + F_n} \tag{13}$$

The F1-score considers both the $P_r$ and $R_e$, and is the harmonic mean of these two values.

$$F1 = 2 \times \frac{Pr \times Re}{Pr + Re} \tag{14}$$

Last, the general average precision is calculated as:

$$G_{Ave} = \frac{Total\ number\ of\ correct\ prediction}{Total\ number\ of\ samples} \tag{15}$$

The model performances are shown in Table 1 and Table 2 below for CranData and Brain4Cars dataset. The models are evaluated based on the six seconds sequential data, which is collected 3.5 seconds before the maneuvers. The models are trained and tested for five times with randomly collected 80% data for training and 20% for testing at each time to generate the mean and STD of the proposed methods. The proposed encoder-decoder CRNN models achieved the state of are results on the two datasets that the lane change intent can be predicted 3.5s before the maneuver with around 90% accuracy. Moreover, it shows that by integrating extra features from a different functional module (lane detection module (**L**) in this case), the model performance can be increased (**CRNN-M-L and CRNN-IR-L**). If extra features are available for the task, the encoder part does not need to be modified, and only the decoder for the specific task has to be updated. Hence, the framework is extendable and allows fast learning and adjustment.

TABLE 1
Results comparison on CranData dataset with a prediction made at 3.5s prior to the maneuver based five separate testing results

| Algorithms | Left Lane Change | | | Right Lane Change | | | Keep Straight | | | General Ave (%) |
|---|---|---|---|---|---|---|---|---|---|---|
| | $Pr$ (%) | $Re$ (%) | $F1$ Score | $Pr$ (%) | $Re$ (%) | $F1$ Score | $Pr$ (%) | $Re$ (%) | $F1$ Score | |
| SVM | 86.2±5.7 | 91.8±4.0 | 88.8±3.5 | 75.5±1.3 | 76.2±3.7 | 75.2±7.5 | 90.3±6.3 | 82.6±8.5 | 86.0±5.3 | 83.6±4.2 |
| HMM | 67.4±5.5 | 81.9±10.9 | 73.7±6.1 | 70.9±7.2 | 86.4±5.6 | 77.7±4.8 | 72.0±8.6 | 39.0±9.2 | 50.3±9.1 | 69.4±5.3 |
| HF-LSTM | 88.5±13.2 | 92.6±6.0 | 89.9±7.2 | 82.9±8.6 | 90.5±8.2 | 86.1±5.7 | 87.0±13.6 | 73.9±14.2 | 78.4±8.0 | 85.6±5.7 |
| CRNN-M | 87.6±8.1 | 89.6±12.6 | 88.1±7.7 | 89.8±11.2 | 92.2±9.3 | 90.7±9.4 | 92.4±4.4 | 85.1±13.6 | 88.0±7.7 | 89.7±6.7 |
| CRNN-IR | 87.1±10.2 | **92.8±4.5** | 89.7±6.7 | 92.5±5.3 | **93.5±4.3** | 92.9±2.4 | **98.2±4.1** | 92.2±7.8 | 94.8±3.6 | 92.7±3.4 |
| CRNN-M-L | 89.0±8.1 | 89.1±3.5 | 88.8±3.9 | 92.6±2.9 | 92.2±8.4 | 92.2±3.8 | 93.4±7.1 | 94.5±8.7 | 93.5±3.4 | 91.7±2.2 |
| CRNN-IR-L | **94.2±6.1** | 88.2±13.9 | **90.8±9.6** | **94.4±5.7** | 92.5±6.1 | **93.3±4.1** | 92.7±7.7 | **98.9±2.3** | **95.6±5.1** | **93.7±5.1** |

TABLE 2
Results comparison on Brain4Cars dataset with a prediction made at 3.5s prior to the maneuver based five separate testing results

| Algorithms | Left Lane Change | | | Right Lane Change | | | Keep Straight | | | General Ave (%) |
|---|---|---|---|---|---|---|---|---|---|---|
| | $Pr$ (%) | $Re$ (%) | $F1$ Score | $Pr$ (%) | $Re$ (%) | $F1$ Score | $Pr$ (%) | $Re$ (%) | $F1$ Score | |
| SVM | 71.8±11.6 | 77.6±5.8 | 74.1±7.3 | 83.7±5.0 | 76.1±11.8 | 79.1±5.4 | 63.9±7.4 | 64.3±15.5 | 63.1±8.4 | 73.8±5.9 |
| HMM | 76.9±10.6 | 92.8±5.0 | 83.8±7.3 | 79.7±10.0 | 91.6±6.0 | 85.1±8.2 | 87.7±7.3 | 61±17.8 | 71±14.9 | 80.4±8.3 |
| HF-LSTM | 84.4±8.9 | 90.1±7.3 | 86.7±2.9 | 86.9±2.0 | 85.4±10.3 | 85.8±4.8 | 84.4±8.3 | 81.3±4.8 | 82.8±5.9 | 84.9±3.7 |
| CRNN-M | 87.2±9.2 | 90.8±3.2 | 88.7±4.7 | 87.8±10.8 | 93.5±6.6 | 89.9±3.7 | 89.4±10.3 | 81.1±9.8 | 84.3±4.0 | 88.2±1.7 |
| CRNN-IR | **92.8±4.7** | **93.2±8.9** | 84.3±.84 | 90.5±7.4 | **95.7±4.0** | 90.2±2.7 | 85.6±12.8 | **87.6±6.8** | 86.2±8.2 | 89.4±4.4 |
| CRNN-M-L | 87.4±7.5 | 89.0±3.7 | 88.1±4.6 | 88.4±10.2 | 94.9±4.9 | 91.3±7.2 | 88.4±6.4 | 81.5±7.2 | 84.6±5.4 | 88.6±2.7 |
| CRNN-IR-L | 86.3±8.0 | 92.9±9.1 | **88.9±2.3** | **91.5±6.5** | 94.2±4.1 | **92.7±3.4** | **91.9±6.4** | 82.1±7.5 | **86.4±3.6** | **89.4±2.2** |

The model classification performance on the three intents on the two datasets is shown in Fig. 5. It shows that lane-keeping intent achieved the most accurate results on the CranData. In contrast, lane change right intent is the most accurate one on the Brain4Cars dataset, which shows the advantage of the proposed method over the results given in [45]. The deeper Inception-ResNet generates slightly better results than the MobileNet. The model training performances are shown in Fig. 6. The intention inference decoder with the high-level features from the two models can be converged after 200 epochs on both of the two datasets.

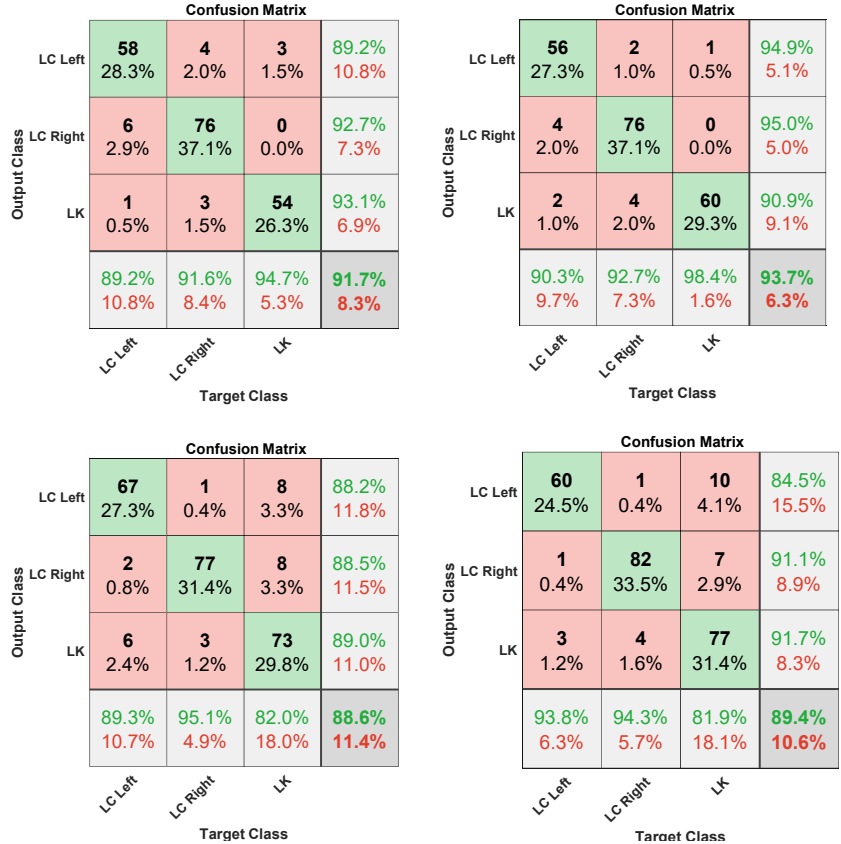

Fig. 5. Confusion matrix and model performance for the intention inference task. The upper two graphs indicate the classification results of MobileNet and Inception-ResNet on the testing data set of CranData. The bottom two graphs show the classification results on the testing data set of Brain4Cars.

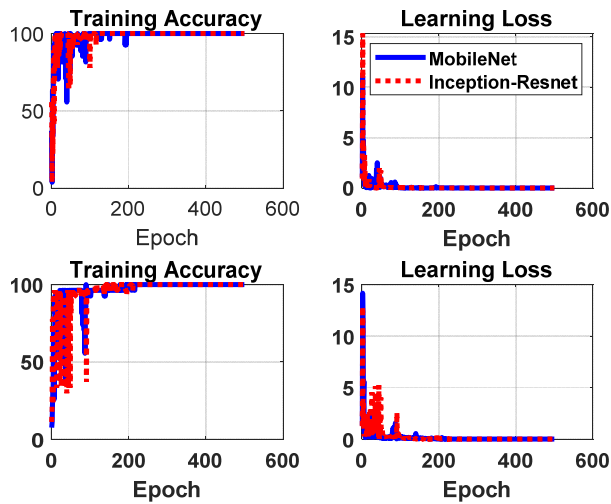

Fig. 6. Intention inference model learning process on the two datasets. The upper two graphs are the learning accuracy and training loss on CranData, and the bottom two are the results on the Brain4Cars dataset.

## C. Driver Emotion Evaluation

In this part, the model performance of the driver emotion recognition task is evaluated. The mental emotion is viewed as a cognitive process that can last for a few seconds and not varied rapidly [52]. Dues to the naturalistic data limitation, only two states, namely, the emotional state and neutral state, are classified. Most of the emotional states in this study are happy states with a small number of surprising cases. The temporal emotional process is similar to the intention process in the last part and can be recognized with similar methods. The HMM and LSTM models are also used as the baseline methods. As the hand-craft head pose and eye gaze features in the last part cannot be used to precisely estimate the facial expression. Two popular features for facial expression and emotion representation are adopted, which are hybrid Histogram of Oriented Gradients (HOG) and the 68 points of facial landmarks [53][54]. A cell size of 32 is chosen for the HOG extractor, and the coordinates of the facial landmarks points in the image plane are used as the features vector. These two feature extraction methods will be combined with the HMM and LSTM to generate four baseline methods. Results comparison between different models on the two datasets is shown in Table 3, and Table 4. respectively.

TABLE 3
RESULTS COMPARISON ON CRANDATA FOR EMOTION CLASSIFICATION BASED ON FIVE SEPARATE TESTING RESULTS

| Algorithms | Neutral Driving | | | Emotional Driving | | | General |
|---|---|---|---|---|---|---|---|
| | $Pr$ (%) | $Re$ (%) | $F1$ Score | $Pr$ (%) | $Re$ (%) | $F1$ Score | Ave (%) |
| HOG-HMM | 80.8±1.7 | 87.3±7.8 | 83.8±4.4 | 47.9±17.3 | 32.0±4.5 | 37.3±5.8 | 74.4±5.9 |
| Facial-HMM | 71.3±13.8 | 72.7±19.3 | 70.1±12.7 | 72.7±19.5 | 66.0±20.7 | 66.5±14.0 | 69.5±12.4 |
| HOG -LSTM | 79.0±9.8 | 85.3±3.8 | 81.7±5.6 | 34.5±12.0 | 29.7±22.2 | 30.4±17.0 | 71.2±8.2 |
| Facial -LSTM | 55.3±2.2 | 92.3±13.3 | 66.1±14.2 | 83.3±28.8 | 22.9±13.5 | 28.1±19.2 | 51.6±14.1 |
| CRNN-M | **96.9±2.0** | 97.4±4.2 | **97.1±2. 2** | **93.3±10.9** | **88.7±9. 2** | **90.4±6.4** | **95.6±5.8** |
| CRNN-IR | 94.6±4.2 | **97.5±2.5** | 96.0±3.2 | 88.9±13.6 | 79.8±18.5 | 83.3±16.0 | 93.7±9.7 |

TABLE 4
RESULTS COMPARISON ON BRAIN4CAR FOR EMOTION CLASSIFICATION BASED ON FIVE SEPARATE TESTING RESULTS

| Algorithms | Neutral Driving | | | Emotional Driving | | | General |
|---|---|---|---|---|---|---|---|
| | $Pr$ (%) | $Re$ (%) | $F1$ Score | $Pr$ (%) | $Re$ (%) | $F1$ Score | Ave (%) |
| HOG-HMM | 91.7±2.9 | 96.3±2.7 | 93.9±1.5 | 78.2±12.9 | 57.8±16.5 | 64.9±11.9 | 89.6±2.6 |
| Facial-HMM | 71.9±5.7 | 4.9±12.7 | 57.2±8.0 | 61.5±3.9 | 80.0±9.3 | 69.1±2.6 | 64.4±3.0 |
| HOG -LSTM | 91.9±4.3 | 92.8±3.3 | 92.3±2.6 | 58.6±11.4 | 56.2±18.0 | 56.0±10.6 | 86.9±4.2 |
| Facial -LSTM | 70.1±14.1 | 54.7±13.4 | 60.8±11.6 | 56.4±15.7 | 72.0±14.1 | 62.5±13.2 | 62.5±10.8 |
| CRNN-M | **99.1±2.0** | 98.6±1.3 | **98.8±0.8** | 92.1±7.4 | **93.3±14.9** | **91.8±7.3** | **98.0±5.6** |
| CRNN-IR | 98.2±2.9 | **99.0±1.3** | 98.6±1.5 | **94.9±7.0** | 88.6±18.6 | 90.5±11.5 | 97.6±7.2 |

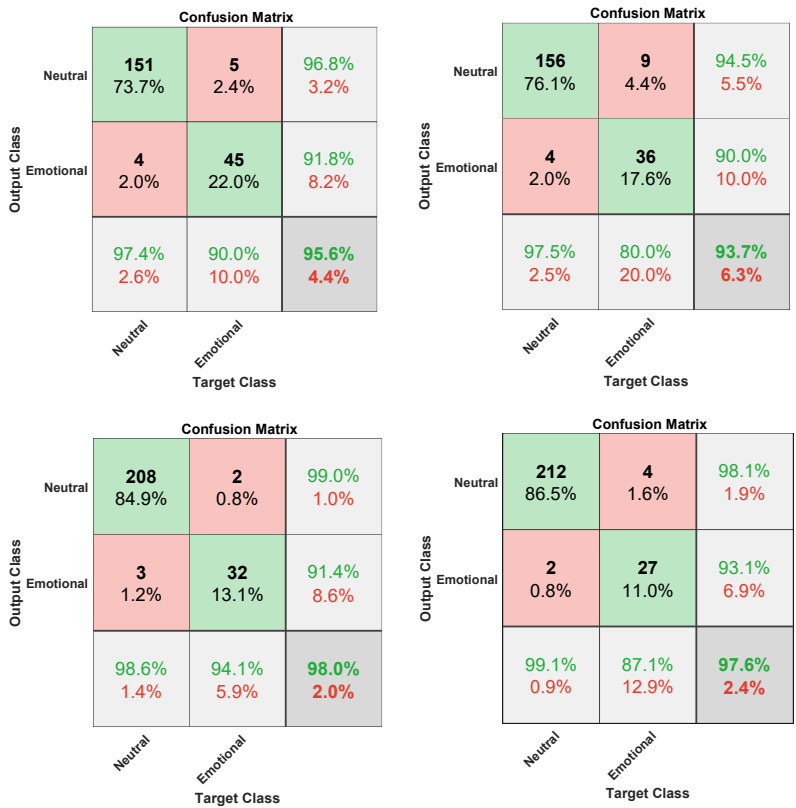

Fig. 7. Confusion matrix and model performance for the emotion recognition task. The upper two graphs indicate the classification results of MobileNet and Inception-ResNet on the testing data set of CranData. The bottom two graphs show the classification results on the testing data set of Brain4Cars.

As shown in Table 3 and Table 4, the proposed encoder-decoder method achieved the most accurate recognition results on the two datasets. The MobileNet achieved slightly better accuracy than the Inception-ResNet in this case. The facial landmarks features are less representative and efficient than the other feature. Regarding the two emotion states, a neutral driving emotion process can be accurately detected. In contrast, the recall (sensitivity) performance on the emotional driving tasks is less precise, which may due to the imbalanced dataset. This can be found in the confusion matrix comparison that is shown in Fig. 7. As given in Fig. 7, the amount of the labeled emotional process is much fewer than the neutral driving samples. The confusion matrix is calculated based on the summation of the five-testing process. The MobileNet shows more precise results than the Inception-ResNet on the two datasets. The model training performance of two different models on the two datasets is shown in Fig. 8. It shows that the binary emotional and neutral states classification encoder is easier to be trained than the intention task and can be converged after 200 epochs on both of two datasets.

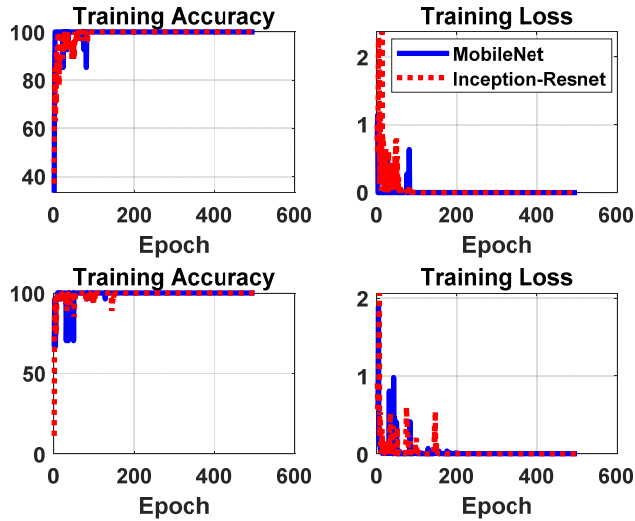

Fig. 8. Emotion inference model learning process on the two datasets. The upper two graphs are the learning accuracy and training loss on CranData, and the bottom two are the results on Brain4Cars dataset.

### D. Analysis of Driver Behaviors, Intention, and emotion

The relationship between driver intent and emotion are analyzed in this part to exploit the connection between these two behaviors. The two behaviors are analyzed from two aspects. First, the proportion of emotional process with respect to the three driving intent are statistically analyzed. Then, based on the driver behaviors and facial expression labels, the relative time intervals between the mirror checking behaviors and the facial expression duration are studied.

First, the proportion of the emotional process for the three driving intent is represented in Fig. 9. In sum, the proportion of the emotional process in the lane-keeping (LK) cases is dramatically larger than that in the lane changing cases. For the CranData dataset, the emotional process accounts for 24.6% in the lane-keeping cases, while only about 16% of the lane changing preparation processes show the emotional facial expression. For the Brain4Cars dataset, the emotional states during the lane changing preparation process are even smaller, which are only 7.8% for the lane change left, and 10.9% for the lane change right. These results and comparisons are made based on the on-hand datasets. However, both of the two datasets indicates that the emotional states are more easily occur during easy driving tasks, such as lane-keeping maneuver. While for the lane change scenarios, the drivers may need to pay more attention to the driving task, which makes them more focus on the situation-aware and vehicle control, and usually show less emotional expression.

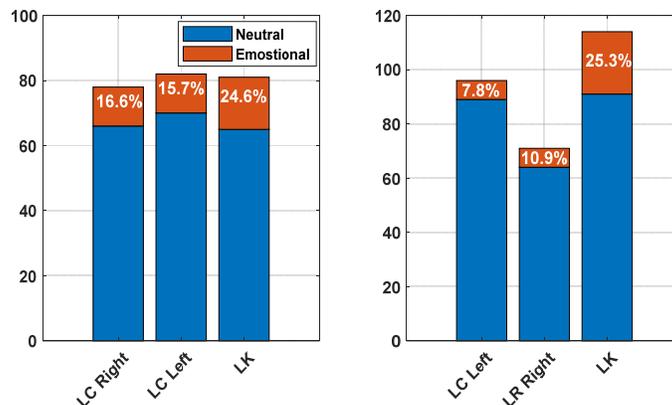

Fig. 9. The proportion of the emotional and neutral process within the three different intent. The left part indicates the statistics on CranData dataset, and the right part shows the statistic results on the Brain4Car dataset.

Next, the relative time intervals between the facial expression and mirror checking behaviors are analyzed. The facial expression represents the driver's emotion, while the mirror checking behaviors show a strong indication for the future driving intention. Hence, analysis of the time intervals between these two states is useful to exploit the dynamics of different mental states. In this step, only the emotional lane change preparation (intended) process is studied. To jointly analyze the emotion and intention state, two critical time intervals are calculated, namely, the time interval between the first facial expression and first mirror checking behavior in the sequence, and the time interval of the last facial expression and las mirror checking moments. The initial time interval and the end time interval are represented as follows.

$$Initial\ Interval = t_e(1) - t_i(1) \tag{16}$$
$$End\ Interval = t_e(end) - t_i(end) \tag{17}$$

where $t_e(1)$ and $t_i(1)$ are the first moments of the facial expression and mirror checking, while $t_e(end)$ and $t_i(end)$ are the last moments of the two behaviors.

Statistical results on the two datasets are illustrated in Fig. 10 below. As shown in Fig. 10, the first facial expression generally occurs about one second earlier than the first mirror checking moment on the two datasets (1.194 s and 1.08 s, respectively). Although, the results of the finish time intervals of the two behaviors are not similar with each, the average finish time intervals are within one second, which means once the mirror checking is finished and the driver is ready to make a lane change maneuver, their obvious facial expressions will not last very long so that the driver can focus on the lane change maneuver.

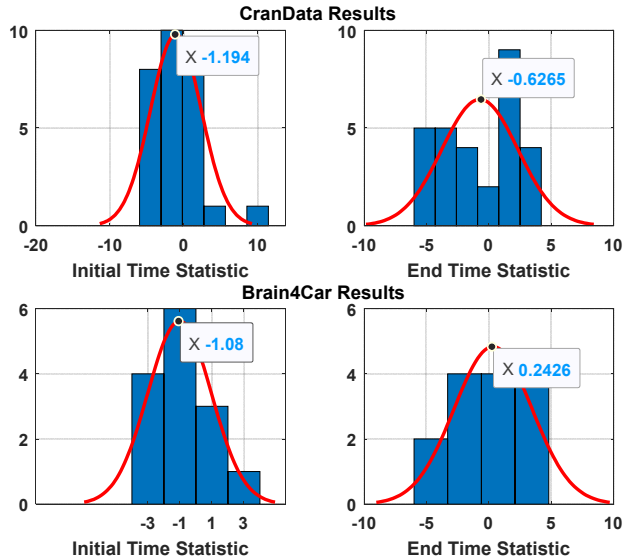

Fig. 10. Statistical analysis of the relative time intervals between the facial expression duration and mirror checking behaviors. Negative values mean emotion start earlier than the intended behavior.

## V. Discussion and Future Works

### A. Advantages and Limitation

In this study, a unified multi-scale driver behavior reasoning framework is proposed and evaluated with three different tasks, which are driver physical behavior recognition, driver intention inference, and driver emotion recognition. Regarding physical behavior recognition, the encoder CNN which was trained with the joint driver behaviors, indicates an efficient representation for driver activity and facial expression. The classification results for the combined driver behaviors are consistent and more accurate than some existing methods [9,15]. Moreover, the model can learn more features and patterns based on jointly labeled driver behavioral data. The visualization for model representation performance shows the CNN encoder can capture the important features for driver behavior classification that particular focus on the facial area. This step is the foundation of multi-task driver behavior reasoning as the following tasks require a proper representation to make driver mental state inference.

Second, two driver mental states are studied, which are driver intention and driver emotion. Driver intention based on the encoder-decoder architecture is compared with multiple existing methods on two datasets. The performance shows the accuracy and efficiency of the proposed methods. In terms of emotion recognition, most of the existing studies focus on the identification of the emotion using statistic images or short image sequences. However, in this study, the emotional state is viewed as a cognitive process like driver intention since the emotional states cannot change rapidly. Based on this assumption, driver emotion is jointly studied with driver intention. It is found that the driver tends to exhibit an emotional driving behavior when the driving tasks are relatively straightforward, such as during normal driving and lane-keeping maneuver. On the contrary, the driver usually keeps a neutral emotion state and focus on the driving task when they generate a lane change intent and prepare for the maneuver.

In sum, the advantages of this study can be summarized as follows.

➢ First, the encoder-decoder framework enables a multi-scale and multi-task driver behavior reasoning. Unlike end-to-end training for multi-tasks learning [38,39], the feature representation process and task inference process are separately trained so that the model is more flexible and easier to be extended.

➢ Second, the feature fusion layer can be used to integrate other features from different modules, and only the decoder part needs to be updated if more function modules are involved. The feature presentation capability of the CNN encoder can be enhanced by introducing more driver state data. This structure is an effective manner for driver behavior reasoning on the intelligent and automated vehicles as driver behaviors data of standard, and single-task are easy to be collected and labeled.

➤ It is hard to estimate the drivers' intention on highly automated vehicles as drivers are usually not driving the vehicle by themselves. However, based on the investigation between driver emotion and driver intention, it is found that driver's emotional states can be used as a clue to estimate whether a driver is concentrating on the road context or not. This can be evidence to determine if the driver is holding a specific driving intention before they take-over the vehicle control authority.

However, there are also limitations exist in this study. One of the primary limitations of this study is limited driver behaviors are collected due to the naturalistic driving task. The emotional states are not very sufficient as it is dangerous to disrupt the driver by sending them too many negative emotional messages or performing secondary tasks such as answering the phone and texting. Hence, driver attention and distraction states are not studied in the current stage, which we think is also essential to the understanding of driver behaviors for both manual driving and autonomous driving tasks.

### B. Future Works

A straightforward task in the future is to exploit and involve more driver behaviors such as driver attention and driver drowsiness, etc. A more comprehensive analysis of the relationship between different driver mental states and cognitive processes is expected to enhance the human understanding intelligence of vehicle automation. In this study, driver activity and facial expression are jointly labeled due to the limitation of data variance. However, in the future, more data can be collected in a distributed manner to improve the generalization ability of the CNN encoder. For example, more challenge and dagerous behavioral data such as more emotions (sad, anger, surprise, tec.), secondary tasks, and fatigue behavior can be collected on the driving simulators or from open public datasets so that the CNN encoder can distinguish more basic driver states [55]. However, increase the data variance and volume with a distributed manner would generate a new consideration, which is the domain adaptation [56]. Introducing more data from other environments should not decrease the original performance on behavior classification. Therefore, domain adaptation needs to be analyzed and evaluated before more data can be adopted.

## VI. Conclusion

In this study, a unified multi-scale driver behavior reasoning framework is proposed. Three different driver behaviors are studied, which are driver physical behaviors (normal driving activities and facial expression), driver intention, and driver emotion. An encoder-decoder multi-task model is proposed based on the integration of CNN and RNN models. By fine-tuning the CNN encoder with the driver's physical behavioral classification task, CNN can learn a precise representation for different driver states. Two pre-trained CNN models, namely MobilyNet V2 and Inception-ResNet V2 are used for the base structure of CNN encoder. The models are evaluated on two different datasets (CranData and Brian4Cars) for highway and urban road driving behavior recognition. Driver behavior classification achieved 95% on the eight defined driver states. The model performance on driver intention inference achieved state-of-the-art results on the two datasets (around 90% in general). Last, the emotional driving process recognition achieved over 95% accuracy. The MobileNet encoder can generate similar results with the deeper network, which makes the framework portable to the on-board embedded systems. The framework is flexible and extendable to be implemented on the intelligent and autonomous vehicles for comprehensive driver/passenger understanding.

## ACKNOWLEDGEMENT

This work was supported in part by the SUG-NAP Grant (No. M4082268.050) of Nanyang Technological University, Singapore, and the Intra-Create Seed Collaboration Grant Project (NRF2019-ITS005-0011), Singapore.